\documentclass[prb,showpacs,twocolumn]{revtex4}%

\usepackage{amsfonts}
\usepackage{amsmath}
\usepackage{amssymb}
\usepackage{graphicx}%
\begin{document}
\title{Conversed spin Hall conductance in two dimensional electron gas in a
perpendicular magnetic field}
\author{Zhigang Wang and Ping Zhang}
\affiliation{Institute of Applied Physics and Computational Mathematics, P.O. Box 8009,
Beijing 100088, P.R. China}

\begin{abstract}
Using the microscopic theory of the conserved spin current [Phys. Rev. Lett.
\textbf{96}, 076604 (2006)], we investigate the spin Hall effect in the two
dimensional electron gas system with a perpendicular magnetic field. The spin
Hall conductance $\sigma_{\mu\nu}^{s}$ as a response to the electric field
consists of two parts, i.e., the conventional part $\sigma_{\mu\nu}^{s0}$ and
the spin torque dipole correction $\sigma_{\mu\nu}^{s\tau}$. It is shown that
the spin-orbit coupling competes with Zeeman splitting by introducing
additional degeneracies between different Landau levels at certain values of
magnetic field. These degeneracies, if occurring at the Fermi level, turn to
give rise to resonances in both $\sigma_{\mu\nu}^{s0}$ and $\sigma_{\mu\nu
}^{s\tau}$ in spin Hall conductance. Remarkably, both of these two components
have the same sign in the wide range of variation in the magnetic field, which
result in an overall enhancement of the total spin Hall current. In
particular, the magnitude of $\sigma_{\mu\nu}^{s\tau}$ is much larger than
that of $\sigma_{\mu\nu}^{s0}$ around the resonance.

\end{abstract}
\maketitle

Spintronics, which combines the basic quantum mechanics of coherent spin
dynamics and technological applications in information processing and storage
devices\cite{Wolf,Awsch,Das}, has grown up to become a very active field in
condensed matter physics. One central issue in spintronics is on how to
generate and manipulate spin current as well as to exploit its various
effects. In the ideal situation where spin is a good quantum number, spin
current is simply defined as the difference between the currents of electron
carried by spin-up and spin-down states. This concept of the spin current has
served well in early studies of spin-dependent transport in metals and
although the ubiquitous presence of spin-orbit coupling inevitably makes the
spin non-conserved, this treatment on spin current sustains to be reasonable
if the net effect of spin-orbit coupling is only considered as one source to
spin relaxation. In recent years, however, it has been found that the
extrinsic or intrinsic spin-orbit coupling can provide a route to generate
transverse spin current in ferromagnetic metals\cite{Hirsch,Zhang} or
semiconductor paramagnets\cite{Dya,Muk1,Sinova} by the driving of an electric
field. The fundamental question of how to define the spin current properly in
the general situation then needs to be
answered\cite{Shi,Jin2005,Murakami2004,Zhang2005,Sun2005,Wang2006}. In most of
previous studies of bulk spin transport, it has been conventional to define
the spin current simply as a combined thermodynamic and quantum-mechanical
average over the symmetric product of spin and velocity operators.
Unfortunately, the recent spin-accumulation experiments\cite{Kato,Wund,Sih}
cannot directly verify it since there is no deterministic relation between
this spin current and the boundary spin accumulation.

Recently, Shi et al.\cite{Shi} have established an alternative definition of
spin current, which is given by the time derivative of the spin displacement
(product of spin and position observable). This differs from the conventional
definition by inclusion of one additional spin torque dipole term. As a
result, the new spin transport coefficients $\sigma_{\mu\nu}^{s}$ have been
shown to consist of two parts, i.e., the conventional part $\sigma_{\mu\nu
}^{s0}$ and the spin torque dipole correction part $\sigma_{\mu\nu}^{s\tau}$.
As one key consequence, the Onsager relation between $\sigma_{\mu\nu}^{s}$ and
other kinds of force-driven transport coefficients has been
shown\cite{Shi,PZhang}. A general Kubo formula for the spin transport
coefficients $\sigma_{\mu\nu}^{s}$ in terms of single-particle Bloch states
has been given\cite{Shi,Ping2007} and applied to study the conserved spin Hall
conductivity (SHC) in the two dimensional hole gas (2DHG). Also, the SHC based
on the new spin current has been recently calculated\cite{Sugimoto} for two
dimensional electron gas (2DEG) with $k$-linear or $k$-cubic spin-orbit
coupling and found to depend explicitly on the scattering potentials.

In this paper we study the conserved spin Hall effect of 2DEG with Rashba
spin-orbit coupling in a perpendicular magnetic field. This system with the
same setup has been studied recently by Shen et al.\cite{Shen} within the
conventional spin current framework. They have found that the conventional SHC
can be made resonant or even divergent by turning the sample parameters and/or
magnetic field $B$. The behavior of the spin torque dipole contribution, and
subsequently the total conserved SHC remains yet to be exploited. As we will
show\ in this paper, the torque term in the SHC can also be made resonant or
divergent by tuning the magnetic field or the Rashba spin splitting. Moreover,
we find that the resonant amplitude of torque term $\sigma_{\mu\nu}^{s\tau}$
is even more prominent than that of the conventional term $\sigma_{\mu\nu
}^{s0}$. The oscillations and resonances in both of these two terms stem from
energy crossing of different Landau levels near the Fermi level due to the
competition between Zeeman energy splitting and spin-orbit coupling. Another
fact we find is that different from the case without magnetic field, the spin
Hall currents contributed from the conventional part and the spin torque
dipole part flow in the same direction, which means an overall enhancement of
the total conserved spin Hall current compared to the conventional spin Hall current.

\begin{figure}[ptb]
\begin{center}
\includegraphics[width=0.50\linewidth]{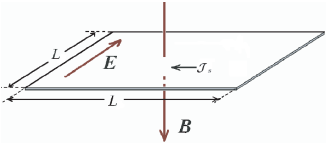}
\end{center}
\caption{(Color online). Illustration of the two-dimensional electron system
in the combined magnetic and electric fields.}%
\end{figure}

We consider a 2DEG with the Rashba coupling in the $x$-$y$ plane of an area
$L\times L$ subject to a perpendicular magnetic filed $\mathbf{B}=-B\hat{z}$.
The electrons are confined between $-L/2$ and $L/2$ in the $y$ direction by an
infinite potential wall, and its wave function is periodic along the $x$
direction. We choose the Landau gauge $\mathbf{A}=xB\hat{y}$. The Hamiltonian
for a single electron of spin-$1/2$ with a Rashba coupling is given by
\begin{equation}
H_{0}=\frac{\vec{\Pi}^{2}}{2m}+\frac{\lambda}{\hbar}(\Pi_{x}\sigma_{y}-\Pi
_{y}\sigma_{x})-\frac{1}{2}g_{s}\mu_{B}B\sigma_{z}, \tag{1}\label{E1}%
\end{equation}
where the confining potential is implied. $m$, $(-e)$, and $g_{s}$ are the
electron's effective mass, charge and effective magnetic factor respectively,
$\mu_{B}$ is the Bohr magneton, $\vec{\Pi}=\vec{p}+e\vec{A}/c$ is the kinetic
operator, $\lambda$ is the Rashba coupling, and $\sigma_{\alpha}$ are the
Pauli matrices. In the presence of the Rashba spin-orbit coupling, the
electron momentum $p_{x}=\hbar k$ along the $x$-direction remains to be a good
quantum number, thus in the Hilbert subspace of given $k$, the Hamiltonian can
be written as
\begin{equation}
H_{0}(k)=\hbar\omega\left[  a_{k}^{\dag}a_{k}+\frac{1-g\sigma_{z}}{2}+\sqrt
{2}\eta(ia_{k}\sigma_{-}-ia_{k}^{\dag}\sigma_{+})\right]  , \tag{2}\label{E2}%
\end{equation}
where $\omega=eB/mc$ is the cyclotron frequency, $\eta=\lambda ml_{b}%
/\hbar^{2}$ is the effective Rashba coupling, and $g=g_{s}m/2m_{e}$, with
$m_{e}$ the mass of a free electron and $l_{b}=\sqrt{\hbar c/eB}$ the magnetic
length. $a_{k}=[y+(\hbar k+ip_{y})c/eB]/\sqrt{2}l_{b}$, so that $[a_{k}%
,a_{k}^{\dag}]=1$, $\sigma_{\pm}=(\sigma_{x}\pm i\sigma_{y})/2$ are the Pauli
matrices. The energy spectrum of $H_{0}(k)$ is given by
\begin{equation}
\epsilon_{ns}=\hbar\omega\left(  n+\frac{s}{2}\sqrt{(1-g)^{2}+8n\eta^{2}%
}\right)  \tag{3}\label{E3}%
\end{equation}
with $s=\pm1$, for $n\geq1$; and $s=1$ for $n=0$. The corresponding
two-component eigenstates for $\epsilon_{ns}$ are given by%
\begin{equation}
|nks\rangle=\left(
\begin{array}
[c]{c}%
\cos{\theta_{ns}}|n\rangle_{k}\\
i\sin{\theta_{ns}}|n-1\rangle_{k}%
\end{array}
\right)  , \tag{4}\label{E4}%
\end{equation}
where $|n\rangle_{k}$ is the eigenstate of the $n$th Landau level in the
absence of the Rashba interaction. For $n=0$, ${\theta_{01}}=0$, otherwise for
$n\geq1$, $\tan{\theta_{ns}}=-u_{n}+s\sqrt{1+u_{n}^{2}}$, with $u_{n}%
=(1-g)/\sqrt{8n}\eta$. The eigenstate $|n,k,s\rangle$ has a degeneracy
$N_{\phi}=L^{2}eB/hc$, corresponding to $N_{\phi}$ quantum values of $k$.

Now let us consider a uniform electric field $\mathbf{E}$ applied in the
$y$-direction. The total Hamiltonian in this case is $H=H_{0}+eEy$ in which
the term $eEy$ is usually treated as a small perturbation. Within the
conserved spin current formalism, the spin current is defined as a
time-derivative of the spin displacement operator, i.e., $\hat
{\boldsymbol{\mathcal{J}}}_{s}=\mathrm{d}(\hat{\mathbf{r}}\hat{s}%
_{z})/\mathrm{d}t$, where $\hat{s}_{z}$ is the spin operator for a particular
component ($z$ here, to be specific), and $\hat{\mathbf{r}}$ is the electron
position operator. Compared to the conventional spin current operator, there
has an extra term, $\hat{\mathbf{r}}(\mathrm{d}\hat{s}_{z}/\mathrm{d}t)$ in
$\hat{\boldsymbol{\mathcal{J}}}_{s}$, which accounts the contribution from the
spin torque dipole. As a consequence, the conserved SHC $\sigma_{xy}^{s}$ as a
linear response to the external electric field defined by
$\boldsymbol{\mathcal{J}}_{s,x}=\sigma_{xy}^{s}E$, includes two components,
\begin{equation}
\sigma_{xy}^{s}=\sigma_{xy}^{s0}+\sigma_{xy}^{s\tau}. \tag{5}\label{E5}%
\end{equation}
The first term in Eq. (\ref{E5}) is the usual conventional SHC, which is ready
to be rewritten in the Landau spinor space (consisting of states $|nks\rangle
$) as
\begin{align}
\sigma_{xy}^{s0}  &  =-\frac{e\hbar}{V}\sum_{\{n,s\}\neq\{n^{\prime}%
,s^{\prime}\},k}\left[  f(\epsilon_{ns})-f(\epsilon_{n^{\prime}s^{\prime}%
})\right]  \times\tag{6}\label{E6}\\
&  \frac{\operatorname{Im}\langle nks|\frac{1}{2}\{\hat{v}_{x},\hat{s}%
_{z}\}|n^{\prime}ks^{\prime}\rangle\langle n^{\prime}ks^{\prime}|\hat{v}%
_{y}|nks\rangle}{\left(  \epsilon_{nks}-\epsilon_{n^{\prime}ks^{\prime}%
}\right)  ^{2}+\delta^{2}},\nonumber
\end{align}
where the velocities are given by%
\begin{align}
v_{x}  &  =\frac{\partial H_{0}(k)}{\partial p_{x}}=\frac{p_{x}}{m}%
+y\omega+\frac{\lambda}{\hbar}\sigma_{y}=\frac{\omega l_{b}}{\sqrt{2}}%
(a_{k}^{\dag}+a_{k})+\frac{\lambda}{\hbar}\sigma_{y}\tag{7}\label{E7}\\
v_{y}  &  =\frac{\partial H_{0}(k)}{\partial p_{y}}=\frac{p_{y}}{m}%
-\frac{\lambda}{\hbar}\sigma_{x}=\frac{i\omega l_{b}}{\sqrt{2}}(a_{k}^{\dag
}-a_{k})-\frac{\lambda}{\hbar}\sigma_{x}\nonumber
\end{align}
and $f(\epsilon_{ns})$ is the equilibrium fermi function. The limit of
$\delta\rightarrow0$ in Eq. (\ref{E6}) should be taken at the last step of
calculation, and as a result, there is no intra-band ($\{n,s\}=\{n^{\prime
},s^{\prime}\}$) contribution. The Kubo formula (\ref{E6}) for the
conventional SHC has been used by most of previous investigations. The second
component $\sigma_{xy}^{s\tau}$ in the conserved SHC (5) comes from the
contribution of the spin torque dipole term\cite{Shi}. In the present context,
its expression reads
\begin{align}
\sigma_{xy}^{s\tau}  &  =-\frac{e\hbar}{V}\lim_{q\rightarrow0}\frac{1}{q}%
\sum_{\{n,s\}\neq\{n^{\prime},s^{\prime}\},k}\left[  f(\epsilon_{nks}%
)-f(\epsilon_{n^{\prime}k+qs^{\prime}})\right]  \times\tag{8}\label{E8}\\
&  \frac{\operatorname{Re}[\langle nks|\hat{\tau}(k,q)|n^{\prime}k+qs^{\prime
}\rangle\langle n^{\prime}k+qs^{\prime}|\hat{v}_{y}(k,q)|nks\rangle]}{\left(
\epsilon_{nks}-\epsilon_{n^{\prime}k+qs^{\prime}}\right)  ^{2}+\delta^{2}%
},\nonumber
\end{align}
where $\tau_{s}(k,q)=\frac{1}{2}[\tau_{s}(k)+\tau_{s}(k+q)]$, $v_{y}%
(k,q)=\frac{1}{2}[v_{y}(k)+v_{y}(k+q)]$, with $\tau_{s}=\frac{1}{i\hbar}%
[s_{z},H_{0}]$. Note that to properly calculate $\sigma_{xy}^{s\tau}$ in
practice, all the terms in Eq. (\ref{E8}) with the subscript $k+q$ should be
expanded at $k$ to first order in $q$.

By substitution of the expressions for single-particle velocities $v_{x,y}$
and eigenstates $|nks\rangle$, and after a tedious but straightforward
derivation, we obtain the conventional part of conserved SHC as follows
\begin{equation}
\sigma_{xy}^{s0}=-\frac{e}{4\pi}\sum_{n,s,s^{\prime}}\frac{f(\epsilon
_{ns})-f(\epsilon_{n+1s^{\prime}})}{[(\epsilon_{ns}-\epsilon_{n+1s^{\prime}%
})/\hbar\omega]^{2}}A(n,s,s^{\prime}), \tag{9}\label{E9}%
\end{equation}
where
\begin{align}
A(n,s,s^{\prime})  &  =(n+1)\cos^{2}{\theta_{ns}}\cos^{2}{\theta
_{n+1s^{\prime}}}\tag{10}\label{E10}\\
&  -n\sin^{2}{\theta_{ns}}\sin^{2}{\theta_{n+1s^{\prime}}}\nonumber\\
&  +\frac{\eta}{\sqrt{2}}(\sqrt{n+1}\cos^{2}{\theta_{ns}}\sin{2\theta
_{n+1s^{\prime}}}\nonumber\\
&  -\sqrt{n}\sin^{2}{\theta_{n+1s^{\prime}}}\sin{2\theta_{ns}}).\nonumber
\end{align}
The other component contributed from the spin torque dipole term turns out to
be given by
\begin{align}
\sigma_{xy}^{s\tau}  &  =\frac{2e\eta}{\pi}\sum_{n,s,s^{\prime}}%
\frac{f(\epsilon_{ns})-f(\epsilon_{n+1s^{\prime}})}{[(\varepsilon
_{ns}-\varepsilon_{n+1s^{\prime}})/\hbar\omega]^{2}}B(n,s,s^{\prime}%
)\tag{11}\label{E11}\\
&  +\frac{2e\eta}{\pi}\sum_{n,s\neq s^{\prime},k}\frac{f(\epsilon
_{ns})-f(\epsilon_{ns^{\prime}})}{[(\epsilon_{ns}-\epsilon_{ns^{\prime}%
})/\hbar\omega]^{2}}C(n,s,s^{\prime}),\nonumber
\end{align}
where
\begin{align}
B(n,s,s^{\prime})  &  =[n\cos{\theta_{ns}}\sin{\theta_{n+1,s^{\prime}}%
}\tag{12}\label{E12}\\
&  -(n+1)\sin{\theta_{ns}}\cos{\theta_{n+1,s^{\prime}}}]\nonumber\\
&  \times\lbrack\sqrt{n(n+1)}\cos{\theta_{ns}}\cos{\theta_{n+1s^{\prime}}%
}\nonumber\\
&  +n\sin{\theta_{ns}}\sin{\theta_{n+1s^{\prime}}}\nonumber\\
&  +\sqrt{2n}\eta\cos{\theta_{ns}}\sin{\theta_{n+1s^{\prime}}}]\nonumber
\end{align}
and%
\begin{align}
C(n,s,s^{\prime})  &  =\sin{(\theta_{ns}-\theta_{ns^{\prime}})}\times\lbrack
n^{2}\cos{\theta_{ns}}\cos{\theta_{ns^{\prime}}}\tag{13}\label{E13}\\
&  +(n-1)\sqrt{n(n-1)}\sin{\theta_{ns}}\sin{\theta_{ns^{\prime}}}\nonumber\\
&  +n\sqrt{2n}\eta\cos{\theta_{ns}}\sin{\theta_{ns^{\prime}}}].\nonumber
\end{align}

A similar expression for the conventional component $\sigma_{xy}^{s0}$ has
also been obtained by Shen et al.\cite{Shen}, while the spin torque dipole
term $\sigma_{xy}^{s\tau}$ is derived for the first time in this paper. The
difference between $\sigma_{xy}^{s0}$ and $\sigma_{xy}^{s\tau}$ is obvious.
Note that the first line in Eq. (\ref{E11}) is obtained by expanding
$|n^{\prime}k+qs^{\prime}\rangle$ in Eq. (\ref{E8}) at $k$ to the first order
in $q$ while remaining other quantities to be their values at $q=0$. Whereas
the second line in Eq. (\ref{E11}) is obtained by a linear expansion of
$\langle n^{\prime}k+qs^{\prime}|$ with respect to $q$. The other terms
occurring in Eq. (\ref{E8}) turn out to take no contributions to $\sigma
_{xy}^{s\tau}$ for the present model.

\begin{figure}[ptb]
\begin{center}
\includegraphics[width=0.70\linewidth]{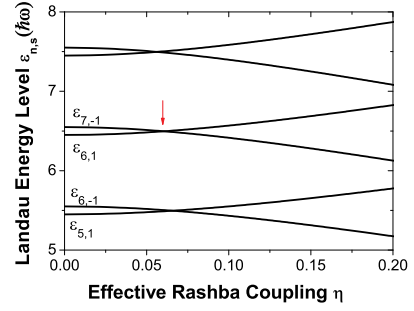}
\end{center}
\caption{Electron Landau levels as a function of effective Rashba coupling
$\eta=\lambda ml_{b}/\hbar^{2}$ for g=0.1.}%
\end{figure}

From Eqs. (\ref{E9}) and (\ref{E11}), specially at zero temperature, one can
see that if $|n,s\rangle$ and $|n+1,s^{\prime}\rangle$ are both occupied or
empty, then the contribution from these two states to the SHC is zero. Thus
only the states near the Fermi level are important to the SHC; this happens in
the situation that the energy-lower state $|n,+1\rangle$ is fully occupied
while the upper state $|n+1,-1\rangle$ is partially occupied or fully empty.
Due to the denominator in Eqs. (\ref{E9}) and (\ref{E11}), one can find that
there will occur resonances at the degeneracies between the Landau levels
$\epsilon_{n,1}$ and $\epsilon_{n+1,-1}$. To show this, we first plot in Fig.
2 the Landau levels as a function of the effective Rashba coefficient $\eta$
for experimentally accessible In$_{0.53}$Ga$_{0.47}$As/In$_{0.52}$Ga$_{0.48}%
$As system\cite{Nitta}. Here the effective magnetic factor is taken to be
$g=g_{s}m/2m_{e}=0.1$. One can see that with increasing the amplitude of the
effective Rashba coefficient $\eta$, the pair of states $|n,s=1\rangle$ and
$|n+1,s^{\prime}=-1\rangle$ approach to cross, implying resonances in
conserved SHC at these level crossings. From expressions for Landau levels
$\epsilon_{n,s}$, one can see that the resonant condition for conserved SHC is
given by
\begin{equation}
\sqrt{(1-g)^{2}+8n\eta^{2}}+\sqrt{(1-g)^{2}+8(n+1)\eta^{2}}=2, \tag{14}%
\label{E14}%
\end{equation}
where $2n\leq\nu<2n+1$, $n=0,1,2,\cdots$, and $\nu$\ is the filling factor
given by the ratio of the total electron number $N_{e}=\sum_{n\mathbf{k}%
s}f(\epsilon_{n\mathbf{k}s})$ to the Landau level degeneracy factor $N_{\phi}%
$, i.e., $\nu=N_{e}/N_{\phi}$. Eq.(\ref{E14}) ensures that the state
$|n,1\rangle$ is fully occupied, while the state $|n+1,-1\rangle$ is not fully
filled. Based on Eqs. (\ref{E9})-(\ref{E13}) we have systematically calculated
the SHC in a wide range of system parameters. \begin{figure}[ptb]
\begin{center}
\includegraphics[width=0.70\linewidth]{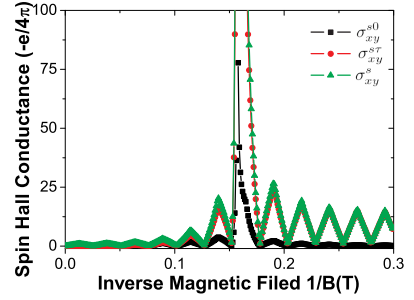}
\end{center}
\caption{(Color online). Conserved SHC and its two components versus $1/B$ at
$T=0$. The parameters are $\lambda=9$ meV nm, $n_{e}=1.9\times10^{-2}/$%
nm$^{2}$, $g_{s}=4$, and $m=0.05m_{e}$, taken for the inversion
heterostructure In$_{0.53}$Ga$_{0.47}$As/In$_{0.52}$Al$_{0.48}$As.}%
\end{figure}Figure 3 plots $\sigma_{xy}^{s}$ and its two components as a
function of inverse of magnetic field (or the effective Rashba coupling
coefficient $\eta$). One can see that there is a pronounced resonance at
filling $\nu=12.6$. At this filling the 13th Landau level (state
$|n=7,-1\rangle$) is partially occupied, while the 12th Landau level (state
$|n=6,1\rangle$) is fully occupied. At the same time these two levels cross at
$\nu=12.6$, as shown in Fig. 2, therefore resulting in a giant resonance
revealed in Fig. 3. Note that the resonant conditions are the same for the
conventional term $\sigma_{xy}^{s0}$ and the spin torque dipole term
$\sigma_{xy}^{s\tau}$, so one can find in Fig. 3 that the resonant peaks in
$\sigma_{xy}^{s}$ and $\sigma_{xy}^{s0}$ have the same location (at $B=6$ T).
Furthermore, it reveals in Fig. 3 that these two conductance components have
the same sign in a wide range of magnetic field, and the resonant amplitude of
$\sigma_{xy}^{s\tau}$ is even more larger than that of $\sigma_{xy}^{s0}$. As
a result the total conserved SHC is prominently enhanced by the inclusion of
the spin torque dipole term in. This result is different from that in the
absence of the magnetic field, wherein it has been verified that the two
components $\sigma_{xy}^{s0}$ and $\sigma_{xy}^{s\tau}$ always compete each
other, giving the opposite contributions to the total spin Hall
current\cite{Ping2007}. Note that the small side peaks occurred in Fig. 3
reflect the variations in filling factor $\nu$ as a function of the magnetic field.

\begin{figure}[ptb]
\begin{center}
\includegraphics[width=0.70\linewidth]{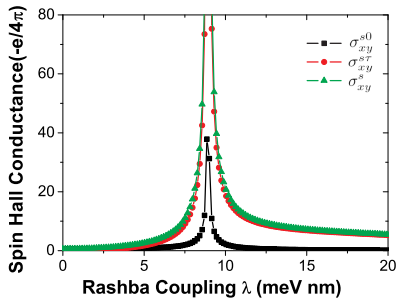}
\end{center}
\caption{(Color online). Conserved SHC and its two components versus Rashba
coupling $\lambda$ at $T=0$. The parameters are $n_{e}=1.9\times10^{-2}%
/$nm$^{2}$, $g_{s}=4$, and $m=0.05m_{e}$.}%
\end{figure}

We also calculate the conserved SHC by only varying the Rashba coefficient
$\lambda$ while the magnetic field remains unchanged. The result is shown in
Fig. 4 for $B=6$ T. The other parameters are set to be the same as used in
Fig. 3. One can see that at $\lambda=9$ meV nm, there occurs a resonance in
both $\sigma_{xy}^{s0}$ and $\sigma_{xy}^{s\tau}$ with different peak
amplitudes and line widths, the resonance behavior is even more prominent for
$\sigma_{\mu\nu}^{s\tau}$. As a consequence, the resonant features of total
conserved SHC is dominated by its spin torque dipole term. The resonance shown
in Fig. 4 corresponds to the same case as Fig. 3 does, i.e., the 12th Landau
level is fully occupied while the 13th Landau level is not fully filled. The
difference is that in Fig. 4 the filling factor $\nu$ remains unchanged when
varying the Rashba coefficient. As a consequence, unlike what is shown Fig. 3,
it reveals in Fig. 4 that there are no side peaks occurred.

In summary, we have studied the spin Hall effect in 2DEG system in a
perpendicular magnetic field by using the conserved definition of spin
current, which includes both the conventional part $\sigma_{\mu\nu}^{s0}$ and
the spin torque dipole correction term $\sigma_{\mu\nu}^{s\tau}$. We have
shown that the conserved SHC can be featured by giant resonances by tuning the
amplitude of system parameters. Furthermore, it has been found that compared
to the previous result of $\sigma_{xy}^{s0}$\cite{Shen}, the resonance
features, including the height and line-width of the resonant peak, are even
more prominent for $\sigma_{xy}^{s\tau}$. It is expected that the present
results could have helpful implications on other aspects involving spin
transport and spin accumulation in the presence of magnetic field.

This work was supported by CNSF under Grant No. 10544004 and 10604010.


\begin{thebibliography}{99}                                                                                               %


\bibitem {Wolf}G.A. Prinz, Science \textbf{282}, 1660 (1998); S.A. Wolf, D. D.
Awschalom, R. A. Buhrman, J. M. Daughton, S. von Moln\'{a}r, M. L. Roukes, A.
Y. Chtchelkanova, and D. M. Treger, Science \textbf{294}, 1488 (2001).

\bibitem {Awsch}\textit{Semiconductor Spintronics and Quantum Computation},
edited by D.D. Awschalom, N. Sarmarth, and D. Loss (Springer-Verlag, Berlin, 2002).

\bibitem {Das}I. \v{Z}uti\'{c}, J. Fabian, and S.D. Sarma, Rev. Mod. Phys.
\textbf{76}, 323 (2004).

\bibitem {Hirsch}J. E. Hirsch, Phys. Rev. Lett. \textbf{83}, 1834 (1999).

\bibitem {Zhang}S. Zhang, Phys. Rev. Lett. \textbf{85}, 393 (2000).

\bibitem {Dya}M.I. Dyakonov and V.I. Perel, JETP \textbf{33}, 1053 (1971).

\bibitem {Muk1}S. Murakami, N. Nagaosa, and S.C. Zhang, Science \textbf{301},
1348 (2003).

\bibitem {Sinova}J. Sinova, D. Culcer, Q. Niu, N.A. Sinitsyn, T. Jungwirth,
and A.H. MacDonald, Phys. Rev. Lett. \textbf{92}, 126603 (2004).

\bibitem {Shi}J. Shi, P. Zhang, D. Xiao, and Q. Niu, Phys. Rev. Lett.
\textbf{96}, 076604 (2006).

\bibitem {Jin2005}P.-Q. Jin, Y.-Q. Li and F.-C. Zhang, e-print, condmat/0502231.

\bibitem {Murakami2004}S. Murakami, N. Nagaosa, and S.-C. Zhang, Phys. Rev. B
\textbf{69}, 235206 (2004).

\bibitem {Zhang2005}S. Zhang and Z. Yang, Phys. Rev. Lett. \textbf{94}, 066602 (2005).

\bibitem {Sun2005}Q. Sun and X.C. Xie, Phys. Rev. B \textbf{72}, 245305 (2005).

\bibitem {Wang2006}Y. Wang, K. Xia, Z.-B. Su, and Z. Ma, Phys. Rev. Lett.
\textbf{96}, 066601 (2006).

\bibitem {Kato}Y. K. Kato, R. C. Myers, A. C. Gossard, and D. D. Awschalom,
Science \textbf{306}, 1910 (2004).

\bibitem {Wund}J. Wunderlich, B. Kaestner, J. Sinova, T. Jungwirth, Phys. Rev.
Lett. \textbf{94}, 047204 (2005).

\bibitem {Sih}V. Sih, R. C. Myers, Y. K. Kato, W. H. Lau, A. C. Gossard and D.
D. Awschalom, Nature Phys. \textbf{1}, 31 (2005).



\bibitem {PZhang}P. Zhang and Q. Niu, e-print, cond-mat/0406436.

\bibitem {Ping2007}P. Zhang Z. Wang, J. Shi, D. Xiao and Q. Niu, e-print, cond-mat/0701293.

\bibitem {Sugimoto}N. Sugimoto, S. Onoda, S. Murakami and N. Nagaosa, Phys.
Rev. B \textbf{73}, 113305 (2006).

\bibitem {Shen}S.-Q. Shen, M. Ma, X.C. Xie, and F.C. Zhang, Phys. Rev. Lett.
\textbf{92}, 256603 (2004).

\bibitem {Nitta}J. Nitta, T. Akazaki, H. Takayanagi, and T. Enoki, Phys. Rev.
Lett. 78, 1335 (1997).
\end{thebibliography}
\end{document}